\documentclass[aps,pra,eqsecnum,twocolumn]{revtex4}
\usepackage{graphicx}
\usepackage{bm}
\usepackage{amsmath,amssymb,amsthm,dsfont,bm}
\usepackage{color}
\usepackage{wasysym}
\usepackage{ulem}
\usepackage{appendix}
\usepackage[dvipsnames]{xcolor}

\usepackage[english]{babel} 
\usepackage[utf8]{inputenc} 
\usepackage{hyperref}
\usepackage{cancel}

\begin{document}
\preprint{}
\title{Eluding Zeno effect via dephasing and detuning}
\author{Julio Cuadrado and Alfredo Luis}
\email{alluis@fis.ucm.es}
\homepage{https://sites.google.com/ucm.es/alfredo/inicio}
\affiliation{Departamento de \'{O}ptica, Facultad de Ciencias F\'{\i}sicas, Universidad Complutense, 28040 Madrid, Spain}
\date{\today}

\begin{abstract}
We analyze some variants of the Zeno effect in which the frequent observation of the population of an intermediate state does not prevent the transition of the system from the initial state to a certain final state. This is achieved by considering system observation involving suitably introduced phase shifts and detunings that leads to a rather rich measurement-induced dynamics by the  alteration of the interference governing quantum evolution. For initial nonclassical states this includes entanglement as a way of evolution from the initial to the final state avoiding the intermediate state. This possibility is presented in a particular physical scenario in the form of a chain of three coupled harmonic oscillators, but we readily show then that the idea can be applied to other physical systems as well, such as atomic-level dynamics. These results are significant for a better knowledge of fundamental quantum concepts as well as regarding suitable applications in the proper control of quantum dynamics, as this is a key feature of modern applications of the quantum theory.

\end{abstract}

\maketitle

\section{Introduction}

The Zeno effect is the alteration of the free dynamics of a system when it is observed. It was initially introduced in the quantum domain as the inhibition of the evolution of a system that is forced by observation to remain in the initial state because of the measurement-induced quantum-state reduction \cite{MS77,IHBW90,PLG14}. Since then, the Zeno effect has been shown to be a dynamical effect decoupled from the reduction postulate \cite{AS94,PN94,VG95,LP96,SP97}, it has also been found in the classical domain \cite{YIK01,PLGS11,GPL15}, and it has been shown that it can not only inhibit evolution, but also accelerate it \cite{LS98,KK00,FGR01,AL02,AL03}, providing in general a suitable subtle way to influence on the dynamics  \cite{FGMPS00,AL01,FP02,FP08,AL11,SHCLCCS14,LS95}.

Much of these results point to a fundamental principle of the evolution of quantum and wavelike systems in general. This is that evolution operates under principles of coherence and interference, as well illustrated by the Huygens-Fresnel principle within the wave theory of light and the Feynman’s path integral approach to quantum mechanics. In both cases the evolution and propagation finally observed arises as the result of interference between multiple allowed paths. According to this picture, we can understand  the Zeno inhibition of dynamics as the result of observation-induced incoherence that prevents constructive interference typical of evolution, while acceleration of the dynamics occurs when decoherence inhibits destructive interference. A nice illustration of this idea can be found in the context of photon emission via spontaneous parametric down-conversion \cite{LP96,LS98,AL02}. 

In this work we intend to advance in the knowledge of the physical mechanisms that operate behind the different versions of the Zeno effect. For this purpose we consider a system which can freely evolve from an initial state $A1$ to a final state $A2$ passing through an intermediate state $B$. We may anticipate that a readily direct observation of whether the system has reached $B$ may lead to the complete inhibition of the evolution, so that the system is frozen in the initial state $A1$. Roughly speaking, we may say that we are blocking all paths from $A1$ to $A2$. But we may as well consider more subtle observations. Inspired by the interference picture of quantum evolution we  may consider system observations altering phase relations between paths and therefore altering the result of the interference. This is the case of considering nonresonant couplings as well as measurement-induced phase shifts. The idea is that this might alter the dynamics so that the system reaches state $A2$ while avoiding the intermediate state $B$.

We illustrate this dynamical effects of observation in a particular physical scenario. This is a chain of three coupled harmonic oscillators. But we also easily show that the idea can be applied to other physical systems. For example, we briefly consider a suitable scheme of two atomic levels $A1$, $A2$ indirectly coupled via a third level $B$ whose population is monitored. 

These results are significant for a better knowledge of fundamental quantum concepts regarding quantum measurement and its dynamical basis and effects. Moreover, this kind of schemes may find suitable applications in the proper control of quantum dynamics, as a key feature of modern applications of the quantum theory \cite{KRJJ23,LLZC23,CP24,TXKG24,HGMDS18,LZW24,BMDMLBWH22,WSQF24,LMFZBHBW23,UL23,RM14}.

\bigskip

\section{Scheme and main goal}

Our scheme is made of a chain of three harmonic oscillators, $A1$, $A2$ and $B$, coupled as illustrated in Fig. 1. For definiteness we will consider them to be three independent modes of the electromagnetic field. Modes $A1$ and $A2$ have the same frequency, being both coupled to $B$, while there is no direct coupling between $A1$ and $A2$, as schematized in Fig. \ref{scheme}. The Hamiltonian of the system, in interaction picture is 
\begin{equation}
    H=\kappa_1 \left( a_1^\dagger b+ a_1 b^\dagger\right)+\kappa_2 \left( a_2^\dagger b+ a_2 b^\dagger\right) - \Delta b^\dagger b ,
    \label{H12}
\end{equation}
where $a_1, a_2$ and $b$  are the corresponding complex-amplitude operators, $\kappa_1$ and $\kappa_2$ are the coupling parameters, and $\Delta$ is the detuning of mode $B$ with respect to modes $A1$, $A2$. 

\begin{figure}[h]
    \includegraphics[width=8cm]{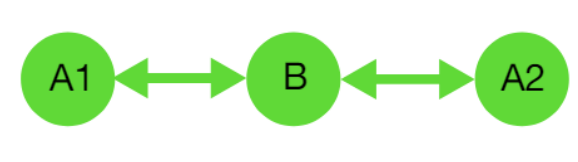}
    \caption{Chain of three coupled harmonic oscillators. Initially only $A1$ is excited. We study the propagation of the excitation form $A1$ to $A2$ depending on the observation of whether the excitation passes through $B$ or not.}
    \label{scheme}
\end{figure}{}

To illustrate the main ideas we will consider a single-photon excitation, initially allocated in mode $A1$. In case of free evolution the photon will evolve to mode $B$ and once there it can evolve to mode $A2$. Then we consider the effect of observation of the population of the intermediate mode $B$. The Zeno logic would say that the frequent observation of whether the photon is in mode $B$ will prevent the transition form $A1$ to $B$ so there will be no chance for the photon to go into the $A2$ mode. 

We will show that this is actually the case in the most simple and standard form of Zeno effect. But we can also show that very simple departures from the standard scenario, simply addding detunig and dephasing, allow the transition from $A1$ to $A2$ even under an arbitrarily precise monitoring of the intermediate mode $B$.    
\bigskip

\section{Observed dynamics}

Free evolution is given by the unitary operator exponential of the Hamiltonian as usual $U = \exp (-i H t)$. Initially, all the light will be concentrated in mode $A1$, while modes $A2$ and $B$ will be initially in their vacuum states $|0 \rangle_2$, $|0 \rangle_b$. Free evolution of duration $\delta t$ will be interrupted periodically at some given times $t_j = j \delta t$ to check whether mode $B$ is in the vacuum state or not. We will consider only those events in which the observation finds the mode $B$ in vacuum. This means that the reduced evolution in modes $A1$ and $A2$ is given by a sequence of nonunitary transformation 
\begin{equation}
\label{nut}
    |\Psi_j \rangle= V_j | \Psi_{j-1} \rangle, \quad V_j =  {}_b \langle 0 | e^{-iH\delta t}| 0 \rangle_b 
\end{equation}
where $|\Psi_j \rangle$ is the state in modes $A1$, $A2$ at times $t_j = j \delta t$, being the initial state $|\Psi_0 \rangle = |\psi \rangle_1 | 0 \rangle_2$, with $ |\psi \rangle$ is in principle an arbitrary state. Later we will include the possibility that the observation introduces a random phase shift at each measurement step. Note that in modes $A1$, $A2$ the initial state is not entangled, although the measurement-induced dynamics may generate entanglement during observed evolution, as we will see in the one-photon case to be considered in Sec. IV. It must be pointed out also that periodicity of observation is not crucial for the scheme since in the appropriate limit of frequent enough observation the results depend just on accumulated phase changes. So we just consider the monitoring to be periodic for the sake of simplicity.

\bigskip

Let us consider a simple mode transformation simplifying calculus. This is a transformation from modes $A1$, $A2$ to some newly defined modes $A$ and $C$ with complex amplitude operators $a$ and $c$ 
\begin{equation}
    a =\cos{\theta} a_1 + \sin{\theta} a_2, \quad c =-\sin{\theta} a_1 + \cos{\theta} a_2,
    \label{12toac}
\end{equation}
such that Hamiltonian (\ref{H12}) reads
\begin{equation}
    H=\kappa \left( a^\dagger b+ a b^\dagger\right) - \Delta b^\dagger b ,
    \label{Hac}
\end{equation}
where
\begin{equation} 
\cos \theta = \frac{\kappa_1}{\kappa}, \; \sin \theta = \frac{\kappa_2}{\kappa}, \quad \kappa = \sqrt{\kappa_1^2 +\kappa_2^2} ,
\end{equation}
and the proper commutations rules are satisfied
\begin{equation}
     \left[a,a^\dagger\right] =\left[c,c^\dagger\right]=1,  \quad  \left[a,c\right]  =\left[a,c^\dagger\right]=0.
    \label{11}
\end{equation}
Besides simplifying calculus, the physical meaning of this transformation is to fully identify the system variables that actually experience the dynamical effects. This turns out to be in the form of the coherent superposition of amplitudes in Eq. (\ref{12toac}). This properly recalls the input-output transformation between the internal and external modes of a two-mode interferometer, so that just the internal mode $A$ is actually coupled to mode $B$. So we may say that this mode definition has to do with the interferometric nature of evolution. Moreover, we provide further physical insight in Sec. IIIA in terms of bright and dark states when presenting an atomic-level realization of this same scheme. 
   
Let us construct the operator $V_j$. We can begin by noting that, after Ref. \cite{LS95}
\begin{equation}
    e^{-iH\delta t} a e^{iH\delta t} = e^{-i\Delta \delta t /2} \left ( \mu^\ast a + \nu^\ast b \right ) ,
    \label{abtrans}
\end{equation}
where 
\begin{equation}
\mu =  \cos (\gamma \delta t ) - i\frac{\Delta}{2\gamma} \sin ( \gamma \delta t ) , \quad \nu = -i \frac{\kappa}{\gamma} \sin (\gamma \delta t ) ,
\end{equation}
being
\begin{equation}
\label{gamma}
    \gamma = \sqrt{\kappa^2 + \Delta^2/4} .
\end{equation}
Note that $\gamma$ has the form of a  Rabi frequency. In Sec. IIIA we will present an atomic-level implementation of this same physical scheme confirming this same interpretation.

In order to suitably derive $V_j$ let us consider the most general form for the pure state $|\Psi_{j-1} \rangle$ in Eq. (\ref{nut}) expressed in the photon-number basis of modes $A1$, $A2$ as
\begin{equation}
    |\Psi_{j-1} \rangle =  \sum_{n_1, n_2}c_{n_1, n_2} a_1^{\dagger n_1} a_2^{\dagger n_2}|0 \rangle_1 |0 \rangle_2 .
\end{equation}
This is, after inverting Eqs. (\ref{12toac}) 
\begin{eqnarray}
  &  |\Psi_{j-1} \rangle =  \sum_{n_1, n_2}c_{n_1, n_2} \left (  
      \cos{\theta} a^\dagger - \sin{\theta} c^\dagger \right )^{n_1} & \nonumber \\
   & \times   \left (  
      \sin{\theta} a^\dagger + \cos{\theta} c^\dagger \right )^{n_2}
        |0 \rangle_1 |0 \rangle_2 , &
\end{eqnarray}
so that after Eq. (\ref{abtrans}) 
\begin{eqnarray}
  &   e^{-iH\delta t} |\Psi_{j-1}\rangle |0 \rangle_b=  \sum_{n_1, n_2}c_{n_1, n_2}  & \nonumber \\ & \times \left [ e^{i\Delta \delta t /2} \cos{\theta}  \left (  \mu a^\dagger + \nu b^\dagger \right ) - \sin{\theta} c^\dagger \right ]^{n_1} & \nonumber \\
   & \times   \left [ e^{i\Delta \delta t /2} \sin{\theta} \left ( 
      \mu a^\dagger + \nu b^\dagger \right ) + \cos{\theta} c^\dagger \right ]^{n_2}
        |0 \rangle_1 |0 \rangle_2 , & \nonumber \\
\end{eqnarray}
and then projecting on the vacuum state in mode $B$ 
\begin{eqnarray}
  &  {}_b \langle 0 | e^{-iH\delta t}| 0 \rangle_b |\Psi_{j-1} \rangle =  \sum_{n_1, n_2}c_{n_1, n_2}  & \nonumber \\ & \times  \left ( 
      e^{i\Delta \delta t /2} \cos{\theta} \mu a^\dagger - \sin{\theta} c^\dagger \right )^{n_1} & \nonumber \\
   & \times   \left (  e^{i\Delta \delta t /2} 
      \sin{\theta} \mu a^\dagger + \cos{\theta} c^\dagger \right )^{n_2}
        |0 \rangle_1 |0 \rangle_2 . &
\end{eqnarray}
With this we can finally obtain $V_j$ in Eq. (\ref{nut}) to be
\begin{equation}
\label{Vj}
  V_j =  \chi_j^{a^\dagger a} , \quad \chi_j = e^{i \phi_j} e^{i \Delta \delta t /2} \mu ,
\end{equation}
and we have already included the possibility that the measurement induces a phase shift $\phi_j$ in mode $A$ as a kind of measurement back action. The final form in Eq. (\ref{Vj}) follows after using that for every function $f(a^\dagger a) $ we have $f(a^\dagger a) a^\dagger = a^\dagger f(a^\dagger a +1)$, and then 
\begin{equation}
    V_j a^\dagger V_j^{-1} =  \chi_j a^\dagger ,
\end{equation}
along with $V_j  |0 \rangle_1 |0 \rangle_2= |0 \rangle_1 |0 \rangle_2$ .

Actually the $V_j$ is a particular element of the Kraus operators defining the backaction of the measurement on the system state in modes $A1$, $A2$. Let us consider that the detection of whether mode $B$ is in vacuum or not is carried out via photon-number detection. In the non referring case, this is ignoring the result of the photo-number measurement in mode $B$, the system state in modes $A1$, $A2$ experiences the following transformation after each measurement $j$ 
\begin{equation}
    \rho_j = \sum_{n=0}^\infty  K_{j,n} \rho_{j-1}  K^\dagger_{j,n} ,
\end{equation}
with the Kraus operators $K^\dagger_{j,n}$ 
\begin{equation}
    K_{j,n} =  {}_b \langle n | e^{-iH\delta t}| 0 \rangle_b ,
\end{equation}
where $\rho_j$ and $\rho_{j-1}$ are the density matrices of the state in  modes $A1$ and $A2$ after and before the measurement, respectively, $| n \rangle_b$ are photon-number states in mode $B$, and $I$ is the identity in modes $A1$ and $A2$. The Kraus operators satisfy the natural constraint  
\begin{equation}
    \sum_{n=0}^\infty  K_{j,n}^\dagger  K_{j,n} = I ,
\end{equation}
and we have that $V_j = K_{j,0}$. 

In this context $\phi_j$ are phenomenological phase shifts that can be introduced during the measurements process by couplings to some unspecified environment, either accidental or introduced on purpose to control the system state in modes $A1$, $A2$. This is introduced just in mode $A$ since this is the only degree of freedom actually coupled to the observation scheme. It is worth noting that all the $\phi_j$ are free variables that in principle can take any value with positive as well as negative signs, being random if they are accidental or deterministic if introduced on purpose. 

\bigskip

Then the complete evolution after $n$ successful measurements checking that mode $B$ is in the vacuum state is 
\begin{equation}
\label{PVj}
     |\Psi_n \rangle = V |\Psi_0 \rangle , 
\end{equation}
with 
\begin{equation}
V=\Pi_{j=1}^n V_j = \chi^{a^\dagger a} ,
\end{equation}
and 
\begin{equation}
\label{chi}
 \chi = \Pi_{j=1}^n \chi_j = e^{i \phi} e^{i \Delta n \delta t /2} \mu^n  ,
\end{equation}
being $\phi$ the accumulated dephasing 
\begin{equation}
    \phi = \sum_{j=1}^n \phi_j .
\end{equation}
Then 
\begin{equation}
\label{VapV}
    V a^\dagger V^{-1} =  \chi a^\dagger ,
\end{equation}
along with $V |0 \rangle_1 |0 \rangle_2= |0 \rangle_1 |0 \rangle_2$ .

\bigskip

Throughout the above analysis the states $ |\Psi_j \rangle$ are not normalized being the norm the success probability, this is the probability that all measurements find the mode $B$ in the vacuum state
\begin{equation}
    P(n,\delta t )= \langle \Psi_n | \Psi_n \rangle =
    \langle \Psi_0 | V^\dagger V |\Psi_0 \rangle .
\end{equation}
    
\bigskip

On what follows we will consider three meaningful cases, all them always in the limit of arbitrarily accurate monitoring of mode $B$, this is with $\delta t = t/n$ for a fixed time interval $t$. These three cases are:

\begin{itemize}
    \item Standard Zeno effect, with no dephasing $\phi =0$ and no detuning $\Delta =0$.

    \item Dephasing $\phi \neq 0$ with no detuning $\Delta =0$.

    \item Detuning $\Delta \neq 0 $ with no dephasing $\phi=0$.
    
    \item Detuning $\Delta \neq 0 $ and dephasing $\phi\neq 0$.
    
\end{itemize}

Finally we note that in our scheme there is no need to distinguish between short and long evolution times. This is because the energy levels involved are discrete, and the measurements are assumed pulsed and instantaneous, so there is no coupling to a continuum of energy levels \cite{DN04}.

\bigskip

\subsection{Three-level equivalence}

The harmonic-oscillator realization we have discussed above is one among many other possible implementations that, reflecting the same physics, may nevertheless differ in their practical possibility of being realized experimentally. In this regard we may consider an alternative implementation in the form of an atomic V level configuration, with a ground state and two excited states coupled to the ground state by two driving lasers, as illustrated in Fig. 2. In this realization the role of modes $A1$, $A2$ and $B$ are played by the atomic states $|A1 \rangle$, $|A2 \rangle$ and $|B \rangle$, respectively. 

The unobserved evolution is governed by the Hamiltonian 
\begin{equation}
    H = \left ( \begin{matrix} 0 & \kappa_1& 0 \\ \kappa_1 &  - \Delta &\kappa_2  \\ 0 & \kappa_2 & 0 \end{matrix} \right ) ,
\end{equation} 
being the matrix basis
\begin{equation}
    | A1 \rangle = \left ( \begin{matrix} 1 \\ 0 \\0 \end{matrix} \right ), \quad
     | B \rangle = \left ( \begin{matrix} 0 \\ 1 \\0 \end{matrix} \right ), \quad
| A2 \rangle = \left ( \begin{matrix} 0 \\ 0 \\ 1 \end{matrix} \right ) . 
\end{equation}
This physical realization shows clearly that Eq. (\ref{gamma}) is actually a Rabi frequency. Moreover, this provides a suitable physical meaning for the corresponding states $|A \rangle$ and $|C \rangle$ within an atomic-coherence picture as the bright state $|A \rangle$, effectively coupled to the ground state, and the dark state $|C \rangle$, fully decoupled from the laser fields driving the transitions \cite{EA96,JM98}. 

In this scheme, the population of level $|B\rangle$ can be readily monitored by resonantly coupling $|B\rangle$ to an auxiliary level $|{\rm aux} \rangle$  that decays spontaneously back to $|B\rangle$  with the emission of photons at a rate $\Lambda$. The presence or absence of the emitted photons reveals that the observed system is in the state $|B\rangle$ or not. So the absence of photons projects the system state in the excited subspace spanned by $|A1 \rangle$, $|A2 \rangle$, being this the counterpart of the projection on the vacuum state in the harmonic oscillator realization above \cite{NSD86,SNBT86,BHIW86,BH96} .

\begin{figure}[h]
    \includegraphics[width=6cm]{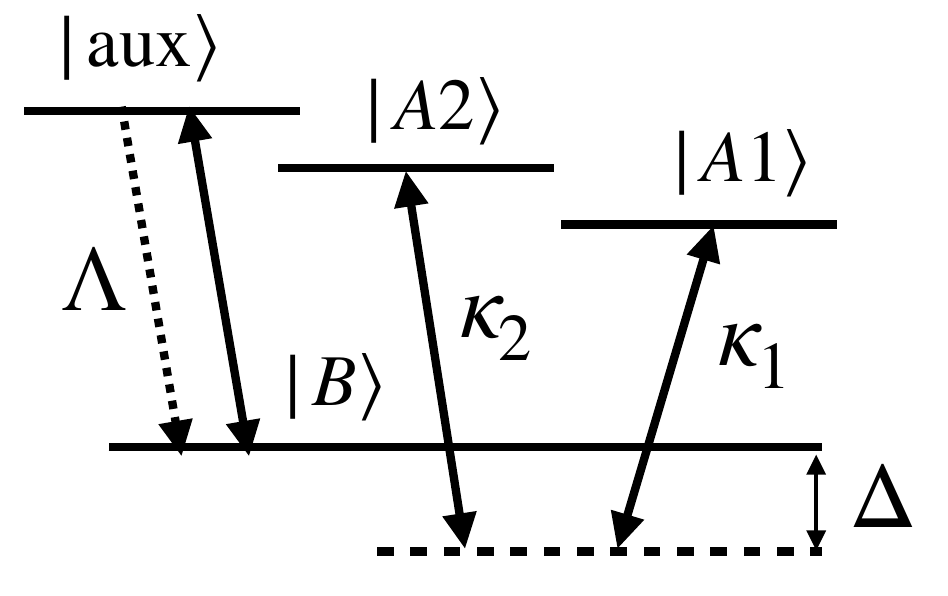}
    \caption{Atomic three-level scheme for another practical realization of the Zeno effect.}
    \label{scheme}
\end{figure}{}

\bigskip

\section{One-photon case}

Let us consider the simple but fully meaningful case of a single photon initially in mode $A1$ at $t=0$, this is 
\begin{equation}
\label{isa1a2}
|\Psi_0 \rangle = | 1 \rangle_1 | 0 \rangle_2 ,
\end{equation}
which is not entangled in modes $A1$, $A2$, while in modes $A$, $C$ the same state is expressed in an entangled form as
\begin{equation}
\label{isac}
|\Psi_0 \rangle = \cos \theta  | 1 \rangle_a | 0 \rangle_c - \sin \theta | 0 \rangle_a | 1 \rangle_c .
\end{equation}

The observed dynamics readily follows after applying Eq. (\ref{PVj}) leading to
\begin{equation}
|\Psi_n \rangle =\chi \cos \theta  | 1 \rangle_a | 0 \rangle_c - \sin \theta | 0 \rangle_a | 1 \rangle_c ,
\end{equation}
that in modes $A1$, $A2$ reads
\begin{equation}
\label{de}
|\Psi_n \rangle =\zeta_1| 1 \rangle_1 | 0 \rangle_2 + \zeta_2 | 0 \rangle_1 | 1 \rangle_2 ,
\end{equation}
with 
\begin{eqnarray}
\label{c1c2}
    & \zeta_1= \chi \cos^2 \theta + \sin^2 \theta , \nonumber \\ & & \\
    & \zeta_2 = \left ( \chi - 1 \right )\cos \theta \sin \theta , & \nonumber
\end{eqnarray}
where $\chi$ is in Eq. (\ref{chi}). We can appreciate how the measurement induced dynamics leads to the appearance of a photon-vacuum entanglement after an initially factorized state.

The probability that all the $n$ measurements find mode $B$ in the vacuum state is 
\begin{equation}
\label{Pndt}
    P(n,\delta t) = |\zeta_1|^2 + |\zeta_2|^2 .
\end{equation}
The main goal is whether the photon can be found in mode $A2$ without having been found in mode $B$. The corresponding conditional probability is 
\begin{equation}
\label{pndt}
    p(n,\delta t) = \frac{|\zeta_2|^2}{|\zeta_1|^2 + |\zeta_2|^2 } .
\end{equation}
We are going to examine these probabilities in the scenarios considered above. 

\subsection{No dephasing $\phi =0$, no detuning $\Delta =0$}

In this case 
\begin{eqnarray}
    & \zeta_1 =  \cos^n (\kappa \delta t )\cos^2 \theta + \sin^2 \theta , \nonumber \\ & & \\
    & \zeta_2 = \left [ \cos^n (\kappa \delta t ) - 1 \right ] \cos \theta \sin \theta .  & \nonumber
\end{eqnarray}
In the limit of arbitrarily accurate monitoring of mode $B$, this is with $\delta t = t/n$ for a fixed time interval $t$ and $n \rightarrow \infty$ we have:
\begin{equation}
    \lim_{n\rightarrow \infty} \cos^n (\kappa t/n) = 1
\end{equation}
so that 
\begin{equation}
    \zeta_1 \rightarrow 1, \qquad \zeta_2 \rightarrow 0 ,
\end{equation}
this is 
\begin{equation}
     P(n,\delta t) \rightarrow 1, \qquad  p(n,\delta t) \rightarrow 0 ,
\end{equation}
and the evolution tends to be completely frozen the photon remaining always in mode $A1$.

\bigskip

\subsection{Dephasing $\phi \neq 0$, no detuning $\Delta =0$}

In this case 
\begin{eqnarray}
    & \zeta_1 =  e^{i \phi} \cos^n (\kappa \delta t )\cos^2 \theta + \sin^2 \theta , \nonumber \\ & & \\
    &\zeta_2  = \left [ e^{i \phi}  \cos^n (\kappa \delta t ) - 1 \right ] \cos \theta \sin \theta .  & \nonumber
\end{eqnarray}
In the same limit of arbitrarily accurate monitoring of mode $B$, this is with $\delta t = t/n$ for a fixed time interval $t$ and $n \rightarrow \infty$ we have:
\begin{eqnarray}
   & \zeta_1 \rightarrow e^{i \phi} \cos^2 \theta + \sin^2 \theta , & \nonumber \\ 
   & & \\
   & \zeta_2 \rightarrow \left (  e^{i \phi}  - 1 \right ) \cos \theta \sin \theta , & \nonumber
\end{eqnarray}
with the following probabilities of success in finding the mode $B$ always in vacuum
\begin{equation}
     P(n,\delta t)  \rightarrow 1, 
\end{equation}
and conditional probability that the photon is found in mode $A2$
\begin{equation}
     p(n,\delta t) \rightarrow 2 \cos^2 \theta \sin^2 \theta \left ( 1 - \cos\phi \right ) .
\end{equation}
In this case we get that the photon is never in mode $B$ while it can be successfully transferred from mode $A1$ to mode $A2$. We can recall that the accumulated dephasing $\phi$ can take any value in principle.

Actually the transfer can be complete $p(n,\delta t) \rightarrow 1$ in the case of $\phi = (2 m+1) \pi$ for integer $m$ and $\kappa_1 = \kappa_2$, this is $\theta= \pi/4$, which is actually the case of a maximally entangled state when expressed in modes $A$ and $C$. This is assuming $\phi$ to be deterministic. If otherwise we consider it as fully random, by averaging over fully random $\phi_j$ we get 
\begin{equation}
    p(n,\delta t)\rightarrow 2\cos^2\theta\sin^2\theta= 2 \frac{\kappa_1^2\kappa_2^2}{\left(\kappa_1^2+\kappa_2^2\right)^2},
\end{equation}
which reaches its maximum value  $p(n,\delta t)\rightarrow 1/2$ for $\kappa_1 = \kappa_2$.

\bigskip

\subsection{Detuning $\Delta \neq 0$, no dephasing $\phi = 0$ }

In order to take full advantage of the detuning let us consider the case of rather strong detuning $\Delta \gg \kappa$ so that 
\begin{equation}
    e^{i \Delta \delta t /2} \mu \simeq e^{- i (\kappa^2/\Delta) \delta t} ,
\end{equation}
and then
\begin{eqnarray}
    & \zeta_1 \simeq e^{- i (\kappa^2/\Delta) n \delta t} \cos^2 \theta + \sin^2 \theta \nonumber \\ & & \\
    &\zeta_2  \simeq \left [  e^{- i (\kappa^2/\Delta) n \delta t} - 1 \right ] \cos \theta \sin \theta .  & \nonumber
\end{eqnarray}
Again in the  limit of arbitrarily accurate monitoring of mode $B$, this is with $\delta t = t/n$ for a fixed time interval $t$, we have that $\zeta_1$ and $\zeta_2$ no longer depend on $n$ and we get the following probabilities of success in finding the mode $B$ always in vacuum
\begin{equation}
     P(n,\delta t) \simeq  1, 
\end{equation}
and the following conditional probability that the photon is found in mode $A2$
\begin{equation}
     p(n,\delta t) \rightarrow 2 \cos^2 \theta \sin^2 \theta \left [ 1 - \cos \left ( \kappa^2 t / \Delta \right ) \right ] . 
\end{equation}
Once again, in this case we get that the photon is never in mode $B$ while it can transferred from mode $A1$ to mode $A2$.

Roughly speaking, in the strong detuning case, the free evolution will never populate the mode $B$, so the measurement has actually no effect. If our system would be just modes $A1$ and $B$ the photon would remain always in the initial mode $A1$. However the coupling of mode $B$ with mode $A2$ allows the migration of the photon from $A1$ to $A2$. This may be pictured as mediated through some virtual intermediate mode different from $B$.

\bigskip

\subsection{Detuning $\Delta \neq 0$ and dephasing $\phi \neq 0$ }

In this case we may readily apply the results of the preceding cases in the very same limits, so that 
\begin{eqnarray}
    & \zeta_1 \simeq e^{i \phi} e^{- i (\kappa^2/\Delta) n \delta t} \cos^2 \theta + \sin^2 \theta , \nonumber \\ & & \\
    &\zeta_2  \simeq \left [  e^{i \phi} e^{- i (\kappa^2/\Delta) n \delta t} - 1 \right ] \cos \theta \sin \theta ,  & \nonumber
\end{eqnarray}
so that
\begin{equation}
     P(n,\delta t) \simeq  1, 
\end{equation}
and
\begin{equation}
     p(n,\delta t) \rightarrow 2 \cos^2 \theta \sin^2 \theta \left [ 1 - \cos \left ( \kappa^2 t / \Delta - \phi \right ) \right ] ,
\end{equation}
obtaining the same conclusions of the preceding case with a different dependence on the measurement induced phases. 

\bigskip

\section{Multi-photon states}

We can easily show that the results obtained for a single photon are reproduced by other initial field states in mode $A1$. As an illustrative example let us comment briefly on the cases of number and Glauber coherent states.

\subsection{Number states}

Let us consider that the initial state in mode $A1$ is a photon-number state:
\begin{equation}
|\Psi_0 \rangle = | N \rangle_1 | 0 \rangle_2 =  \frac{1}{\sqrt{N!}}a^{\dagger N}_1  |0 \rangle_1 |0 \rangle_2 ,
\end{equation}
that in modes $A$ and $C$ reads
\begin{equation}
|\Psi_0 \rangle = \frac{1}{\sqrt{N!}} \left ( \cos{\theta} a^\dagger - \sin{\theta} c^\dagger \right )^N  |0 \rangle_a |0 \rangle_c .
\end{equation}
After Eqs. (\ref{PVj}) and (\ref{VapV}) we get 
\begin{equation}
|\Psi_n \rangle =  \frac{1}{\sqrt{N!}} \left ( \zeta_1 a^\dagger_1 + \zeta_2 a^\dagger_2 \right )^N  |0 \rangle_1 |0 \rangle_2 .
\end{equation}
leading to 
\begin{equation}
 |\Psi_n \rangle =  \sum_{j=0}^N \left ( \begin{matrix} N \\ j \end{matrix}\right )^{1/2} \zeta_1^{N-j} \zeta_2^{j}  |N-j\rangle_1 |j \rangle_2 ,
\end{equation}
where $\zeta_1$, $\zeta_2$ are the same in Eq. (\ref{c1c2}). The probability that the $n$ measurements find the mode $B$ in vacuum is 
\begin{equation}
     P(n,\delta t) = \langle \Psi_n |\Psi_n \rangle = \left ( |\zeta_1 |^2 + |\zeta_2 |^2 \right )^N .
\end{equation}
In this case we can measure the amount of light transferred to the $A2$ by the mean number of photons:
\begin{equation}
    \bar{N_2} = \frac{\langle \Psi_n |a^\dagger_2 a_2 | \Psi_n \rangle }{\langle \Psi_n |\Psi_n \rangle} = N \frac{|\zeta_2|^2}{|\zeta_1|^2 + |\zeta_2|^2 } .
\end{equation}
Therefore, we find these are a simple scaled versions of the one-photon results in Eqs. (\ref{Pndt}) and (\ref{pndt}).

In this multiphotonic case we have examined not the transfer of the initial state as a whole, but the transfer of the amount of light represented by the mean number of photons. Since the number of photons is in principle a random variable we may include as well in the analysis its variance, say 
\begin{equation}
    \Delta^2 N_2= \frac{\langle \Psi_n |\left (a^\dagger_2 a_2 \right )^2| \Psi_n \rangle }{\langle \Psi_n |\Psi_n \rangle} - \bar{N}^2,  
\end{equation}
leading to 
\begin{equation}
       \Delta^2 N_2= N \frac{|\zeta_1|^2 |\zeta_2|^2}{\left (|\zeta_1|^2 + |\zeta_2|^2 \right)^2 } = \frac{\bar{N_1}  \bar{N_2}}{N} .
\end{equation}
So, as the transfer approaches to be complete $|\zeta_1| \rightarrow 0$ and $|\zeta_2| \rightarrow 1$ the uncertainty tends to vanish. 

\bigskip

\subsection{Coherent states}

Let us consider that the initial state in mode $A1$ is a Glauber coherent state:
\begin{equation}
|\Psi_0 \rangle = | \alpha \rangle_1 | 0 \rangle_2 ,
\end{equation}
where $ | \alpha \rangle_1$ is a Glauber coherent state, eigenvector of the complex-amplitude operator, which can be expressed as 
\begin{equation}
|\Psi_0 \rangle = e^{-|\alpha|^2/2} e^{\alpha a_1^\dagger} |0 \rangle_1 |0 \rangle_2 .
\end{equation}
Following the same steps already followed above we find
\begin{equation}
\label{fs}
|\Psi_n \rangle = e^{-|\alpha|^2 [1-|\zeta_1|^2-|\zeta_2|^2]/2} |\alpha \zeta_1 \rangle_1 | \alpha \zeta_2 \rangle_2 ,
\end{equation}
where $|\alpha \zeta_1 \rangle_1 | \alpha \zeta_2 \rangle_2$ are normalized Glauber coherent states with $\zeta_1$, $\zeta_2$ in Eq. (\ref{c1c2}).  Regarding the mean number of photons transferred and its variance we readily get, in accordance with the standard result for coherent states,
\begin{equation}
\label{mn2}
\bar{N_2} = |\alpha |^2 |\zeta_2|^2, \qquad \Delta^2 N_2 = \bar{N_2} ,
\end{equation}
where $|\alpha |^2$ is the mean total number of initial photons. So, all the results found in the one-photon can be equally repeated here in all the observation scenarios considered, again with the same conclusions. 

\bigskip

It can be seen that a series of powers on $\alpha$ of the coherent case actually reproduces the number case, including the one-photon example. In this regard, in the limit of arbitrarily precise observation of mode $B$ we get that $V$ approaches the unitary operator 
\begin{equation}
 V \rightarrow  e^{i \left ( \phi- \kappa^2 t /\Delta \right ) a^\dagger a} ,
\end{equation}
which is a phase shift in mode $A$ and a lossless beam splitter in modes $A1$ and $A2$. So the cases considered in this work, number and coherent, behave alike because of the complex-amplitude operator transformation induced by this effective beam splitter that underlies all the cases examined.

Finally let us point out two interesting results exclusive of the coherent case which are absent in the photon-number example. The first one is that the measurement-induced evolution proceeds without entanglement, as we have always a product state in Eq. (\ref{fs}). The second one, after Eq. (\ref{mn2}) and its equivalent for mode $A1$, is that the total mean number of photons is not conserved as far as $|\zeta_1 |^2 + |\zeta_2 |^2  \neq 1$, even after finding the mode $B$ always in the vacuum state.

\section{Conclusions}

We have analyzed some variants of the Zeno effect in which exhaustive observation of the population of an intermediate state does not prevent the transition of the system from the initial state to a certain final state. The result can be understood as paradoxical since, according to the most standard version of the Zeno effect, the precise observation of the occupation of the intermediate state freezes the system in the initial state so never evolves to the final state. 

In our case the evolution to the final state is allowed in spite of the frequent observation. This is because dephasing and detuning alter the interference that is always behind any quantum or wavelike evolution. We can recall that the different versions of Zeno effect are actually interferometric in nature, because measurement-induced back action impedes constructive or  destructive interference, depending on the context.

As a rather interesting point of these dynamical cases of Zeno effect we have the potential emergence of entanglement during evolution, as it is clear in the one photon case in  Eq. (\ref{de}). Note that although the initial state factorizes in the original modes $A1$ and $A2$, the same field state is actually entangled in modes $A$ and $C$. This mode transformation is the same one induced by a beam splitter, that it is known to produce entanglement when the initial state is nonclassical \cite{VS14}. So we may say that in this case the photon transfer occur via entanglement. Note also that this not need to be always the case, as illustrated by the coherent-state case in Eq. (\ref{fs}), that always factorizes in all modes considered. 

We may also consider the immediate benefit that Zeno dynamics offer. This is the suitable control at demand of the observed-system state, that is certainly continuously certified by the very same observed mechanism that produces the effect. Moreover, we have also seen that this results hold for other physical systems such as three level atomic systems, which can be used in quantum gates design. Hence, this way to control the populations on different levels could also be used to further add functionalities to quantum logic gates.

A very suggestive feature of the cases we have discussed here is that they invoke physical processes that in other contexts mark the transition from quantum to classical physics. For example, decoherence caused by random phases is a known mechanism of emergence of the classical world from the quantum one, something that has already been studied precisely in this same context of the Zeno effect \cite{WHZ03,LAL20}. On the other hand, the lack of resonance guarantees, for example, that in the interaction of light with matter the classical Lorentz model of the atom is perfectly valid. 


\end{document}